\documentstyle[12pt]{article}
\begin{document}
\newcommand\Ds\displaystyle
\def\bm{\boldmath }
\def\k{\kappa }
\def\p{\partial }
\def\l{\lambda }
\def\L{\Lambda }
\def\s{\sigma }
\def\o{\omega }
\def\O{\Omega }
\def\a{\alpha }
\def\e{\equiv }
\def\b{\beta }
\def\g{\gamma }
\def\G{\Gamma }
\def\fr{\frac }
\def\ra{\rightarrow }
\def\d{\delta }
\def\D{\Delta }
\def\t{\tau }
\def\h{\frac{1}{2}}
\def\ov{\overline }
\title{On the problem of  unboundedness from below of the spinor QED
Hamiltonian}
\author{L.G.Zastavenko\\
Laboratory of Theoretical Physics\\Joint Institute for Nuclear Research\\
 \\141980 Dubna Russia }
\maketitle
\begin{abstract}
We show that  the Hamiltonian $ h= H_{QED}+H_2$, where $H_{QED}$ is
the spinor QED Hamiltonian and $H_2$ is the positive transversal photon
mass term, is  unbounded from below if the electromagnetic coupling constant
$e^2$ is small enough,  $e^2<e^2_0 $, and the transversal photon squared
mass parameter $M^2$ is not too large: $0\leq M^2<e^2l^2c$, here, $l$ is the
cut-off parameter, and $c$ and $e^2_0$,  positive constants which do not
depend on any parameters.
\end{abstract}
\bigskip

{\bf 0. INTRODUCTION}

\vspace{5mm}

 F.Palumbo [1] showed that the spinor QED Hamiltonian $H_{QED}$
is unbounded from
 below. He  got this interestlng result considering, in fact,  the operator
$$\overline{H}=\frac{\int dS(0)\O^*_{pr}H_{QED}\O_{pr}}
{\int dS(0)\O^*_{pr}\O_{pr}} \eqno(0.1)$$
here $dS(0)\e  \prod_{{\bf k}\neq 0,\l=1,2 }dq({\bf k},\l )$,
and $\O_{pr}$ is a  trial function  depending on
the transversal
photon variables $q({\bf k},\l ), {\bf k}\neq 0, |{\bf k}|<l,
\l =1,2$, on electron and
positron degrees of freedom and on the zero momentum mode
vector potential variable
${\bf q}(0)$ [1],[2].
Thus, $\ov {H}$ depends on ${\bf q}(0)$ and $\p /\p {\bf q}(0)$.
The simplest choice of the trial function enabled F.Palumbo to get the
operator $\ov {H}$  of the form
$$
H_{QED}\ra \ov {H}_1=-\h (\frac{\p }{\p {\bf q}(0)})^2+{\bf q}(0)\cdot {\bf a}.
\eqno(0.2)
$$
Here, ${\bf a}$ does not depend on ${\bf q}(0)$.
The operator $\ov {H}_1$ is obviously unbounded from below.
This result is obliged entirely to the new term $H_p$, see eq. (1.8).
F.Palumbo has shown that {\em this term has to be introduced into the
QED Hamiltonian that is given in  text books}-see, e.g., [3].
In this work I substitute
$$dS({\bf r})\e  d{\bf q}(0)
\prod_{{\bf k}\neq 0,{\bf k}\neq \pm {\bf r},\l =1,2}
dq({\bf k},\l )
\eqno(0.3a)
$$
for $dS(0)$
in eq.(0.1) (here ${\bf r}$ is a fixed value of the photon momentum) and use
a more sophisticated choice
of the trial function (see eqs. (2.1), (2.2), (2.7), (2.8), (2.19) and
(2.23)). The result is the formula
$$
H_{QED}\ra \ov {H}_2=- \sum _{\l =1,2}
[\frac{\p }{\p q({\bf r},\l )}\frac{\p }{\p q(-{\bf r},\l )}
+q({\bf r},\l )q(-{\bf r},\l )\g ^2]
$$
$$+const
+O(1/V),
\eqno(0.3)
$$
$$
\g ^2\e [e^2c(m/l,r/l)+e^4d(m/l,r/l)]l^2.
\eqno(0.4)
$$
Here, $e^2$
is the nonrenormalized coupling constant, $m$ is the electron mass
parameter in the Lagrangian of the spinor QED, $l$ is
the momentum cut-off parameter,
$V$ is a large periodicity volume, $l\ra \infty $ , $V\ra \infty $ ,
and $c(x,y)$ and $d(x,y)$,
some functions, $c(x,0)$ being positive , $c(0,0)>0,$  $d(x,y)$
is bounded if $x\geq 0, y\geq 0$.
Equations (0.3) and (0.4), if $e^2\ll 1$, indicate the existence of the
negative photon squared mass of the order of magnitude $\sim e^2l^2$ and
unboundedness from below of the operators $\ov {H}_2$ and $H_{QED}$,
as well as the operator $h$ of the abstract.

{\bf 0.} The article is organized as follows.
Section {\bf 1.} contains rather
voluminous  preliminary explanations
concerning: a) the spinor QED Hamiltonian and gauge-transformational
properties of the variables, on which this Hamiltonian depends,
b) my method of the cut-off, c) the problem of
competibility of the realistic cut-off with the Lorentz invariance
  and d) the idea of my proof.
Section {\bf 2.} contains the proof of the statement of the abstract,
i.e. the derivation of eqs. (0.3) and (0.4).
Appendix A contains derivation of some formulas which are necessary
for the proof of sec.{\bf 2}.

\vspace{5mm}

{\bf 1. SOME PRELIMINARY EXPLANATIONS}\\
\vspace{5mm}\\
Here, I shall consider the Hamiltonian $h$,
 $$h=H_{QED}+H_2 \eqno(1.1)$$
 which is the sum of the QED cut-off Hamiltonian $H_{QED}$ [1],[4],
 $$ H_{QED}= H_{0ph} + H_{0f}+H_1+H_c+H_p, \eqno(1.2)$$
and the positive term $H_2$:
$$
 H_{0ph} =    \int [({\dot {\bf B}}^{tr})^{2}
+(rot{\bf B}^{tr})^2]d{\bf x}/2  $$
$$ =  \sum_{{\bf k}\neq 0, \lambda =1,2} [-\frac{\partial }
 {\partial q({\bf k},\lambda )}\frac{\partial }{\partial q(-{\bf k},\lambda )}
 +k^2q({\bf k},\lambda )q(-{\bf k},\lambda )]/2,\eqno(1.3)   $$

$$
H_{0f}= \sum E({\bf p})[a^*({\bf p},\sigma )a({\bf p},\sigma )
 +b^*({\bf p},\sigma )b({\bf p},\sigma )] $$
$$ = \int \psi_1^*({\bf x})
[-i{\hat \alpha \nabla }+\beta m]\psi_1({\bf x})d{\bf x},
\eqno(1.4)$$
here $ E({\bf p})=\sqrt {{\bf p}^2+m^2}$ ,
 $$H_1=e\int \psi _1^*({\bf x}){\hat \alpha }\psi _1({\bf x})
{\bf B}^{tr}({\bf x})d{\bf x}
 , \eqno(1.5)$$
 $${\bf B}^{tr}({\bf x})=\sum _{{\bf k}\neq 0, \lambda =1,2}
{\bf e}({\bf k},\lambda )
  q({\bf k},\lambda )e^{i{\bf kx}}/\sqrt{V}
 , \eqno(1.6)$$
 $$H_c=e^2\sum _{{\bf k}\neq 0}\frac{\rho ({\bf k})\rho (-{\bf k})}
{2V{\bf k}^2},\eqno(1.7)$$
here $\rho ({\bf k})= \int \psi _1^*({\bf x})\psi _1({\bf x})
e^{i{\bf kx}}d{\bf x}, $
 $$H_p=-\h(\frac{\partial }{\partial {\bf q}(0)})^2+
 e{\bf q}(0)\int \psi _1^*({\bf x}){\hat \alpha }\psi _1({\bf x})
d{\bf x}/\sqrt{V}
 , \eqno(1.8)                                              $$

 $$H_2=M^2/2\int [{\bf q}(0)/\sqrt{V}+{\bf B}^{tr}({\bf x})]^2
d{\bf x},\eqno(1.9)$$
here, $M^2>0$.
 In equations (1.2)-(1.9) $H_{0ph}$ is the Hamiltonian of free
 transversal photons, ${\bf e}({\bf k},\lambda )$ is the polarization vector
 of a photon with the momentum ${\bf k}$ and polarization index $\lambda $,
$({\bf k}\cdot {\bf e}({\bf k},\l))=0$,
 $({\bf e}^*({\bf k},\lambda 1)\cdot {\bf e}({\bf k},\lambda 2))=
 \delta _{\lambda 1, \lambda 2}, {\bf e}^*({\bf k},\lambda )=
 {\bf e}(-{\bf k},\lambda ),  {\bf q}(0)$ is the spatially independent zero
 momentum mode of the vector potential [1], [2], [4], ${\bf q}(0)=
 \int {\bf B}({\bf x})d^3x/\sqrt{V}$,  $H_{0f}$ is the Hamiltonian of free
 electrons and positrons, $m$ is the fermion mass parameter,
 $a({\bf p},\sigma )$ and $b({\bf p},\sigma )$ are the annihilation operators
 of the electron and positron with  the momentum ${\bf p}$ and spin projection
 $\sigma , \psi _1({\bf x})=\sum [u({\bf p},\sigma )a({\bf p},\sigma )+
 v({\bf p},\sigma )b^*(-{\bf p},\sigma )]e^{i{\bf px}}/\sqrt{V},
u({\bf p},\sigma )$ and
 $v({\bf p},\sigma )$ are the solutions of the Dirac equation with
 the energy $\pm E({\bf p})$. The Hamiltonian $H_1$ describes
 the interaction of photons, electrons and positrons.
 The Hamiltonian $H_c$ describes the Coulomb interaction between
 electrons and positrons. In the Fourier representation of the
 functions ${\bf B}^{tr}({\bf x})$ and $\psi _1({\bf x})$
 one
 has $|{\bf k}|<l, |{\bf p}|<l$, (see, however, items 1.4. and 1.4.1.),
 $l$ being the cut-off parameter, $V$ is large periodicity cube,
 $V$ tends to infinity.

 {\bf 1.} The Hamiltonian $H_{QED}$ is expressed in terms of the gauge
 invariant quantities ${\bf B}^{tr}, {\bf q}(0)$ and $\psi _1$. If the
 functions $A_{\mu}(x)$ and $\psi (x)$  in the Lagrangian
 $$L=-(\partial _{\mu }A_{\nu }-\partial _{\nu }A_{\mu })^2/4
 -\psi ^*\gamma _4[\gamma _{\mu}(\partial _{\mu}-ieA_{\mu})+m]\psi $$
 which gives rise to the Hamiltonian $H_{QED}$, undergo gauge transformation
 $A_{\mu} \rightarrow A_{\mu}+\partial \lambda (x)/\partial x_{\mu},
 \psi \rightarrow e^{ie\lambda (x)}\psi $, the variables ${\bf B }^{tr}
 , {\bf q}(0) $ remain constant, and the function $\psi _1({\bf x})$
 acquires a spatially independent phase multiplier (see, e.g.[4]).
 Thus, my method of the cut-off does not break down the gauge invariance.\\
 $H_p$ is the zero momentum mode term of the QED  Hamiltonian [1].
 The discovery of it enabled F.Palumbo to prove
 the unboundedness of the Hamiltonian $H_{QED}$
 from below [1].\\

 {\bf 1.1.}  Here, I am going to
 show that not only the Hamiltonian $H_{QED}$, but also the Hamiltonian
$h=H_{QED}+H_2$, $H_2$ being positive, see eq.(1.9),
is  unbounded from below if the
 positive quantity $M^2$ is not too large so that the inequality
 $$M^2<\g ^2,\eqno(1.10)$$
 is fulfilled (see  eq.(0.4)).We have $c(m/l,0)>0$,
thus $\g ^2>0$ if $e^2\ll 1$ independently of the sign of the
function $d(m/l,0)$. {\em Thus, for small values of $e^2$ my consideration
gives  stronger result than that  by F.Palumbo.}

 {\bf 1.1.1.} Omitting the Coulomb term in the present consideration, one
 gets an essentially analogous consideration for the massless Yukawa model.
 For this model also there holds the statement analogous to that of
 item 1.1 , where, however, one has to take {d=0}.

 {\bf 1.1.2.} Let us note that  one cannot prove the
 unboundedness from below of the scalar QED-s
Hamiltonian via the method of this work:
in case of the scalar QED
 the squared oscillator frequency in eq.(0.3)
is positive if $e^2\ll 1$(and has the order of magnitude $\sim e^2l^2$) .
 Thus, combining the spinor field with several charged scalar fields,
 one  hopes to construct the QED model whose Hamiltonian is bounded
 from below.

{\bf1.2} The Palumbo's proof of the
unboundedness from below of the QED Hamiltonian
( see eq. (0.2))  is essentially based on using zero momentum mode
${\bf q}(0)$ [1]. On the contrary, my consideration has little to do
with the zero momentum mode term $H_p$. {\em The unboundedness
from below of the spinor $QED$ Hamiltonian (if $e^2\ll 1$)
is the consequence of the fact, that in any QFT model with a
trilinear intraction $g\times fermion^*\times boson \times fermion$
the $g^2$
perturbation theory correction to the boson squared mass is negative}.
The latter fact is common knowledge since long ago.

{\bf 1.3.} It is worth  mentioning here that {\em the starting point of my
 key construction (2.7), (2.8)}  was an attempt to
get a variational estimate of the type  of eq.(0.1)  by
 using the trial
function
$\O_{pr},\\
 \O_{pr}\e exp(\k _0+\k _1e)|0>$ where the function $\O _0 $,
$$\O _0\e exp(\sum_{n=0}^{\infty }\k _ne^n ) |0>\e e^K|0>\eqno(1.11)$$
is the ground state wave function of the Schroedinger equation
$$(H_{QED}-E)\O=0,\eqno(1.12)$$.
and $|0>$ is the state of the bare fermion vacuum.

{\bf 1.3.1.}  The exponential representation (1.11), being substituted
into the Schroedinger equation, enables one to recurrently find
 functions-operators $\k _n, n=0,1,2,...$
 (were the ground state to exist). These functions-operators
depend on the zero momentum mode variables ${\bf q}(0)$, on the photon
variables
 $q({\bf k},\l ), {\bf k}\neq 0$ and on the electron
and positron creation operators. Of course, the
exponential representation is equivalent to the straightforward linear
representation
$$\O _0=\sum_{n\geq 0}\O_{0n}e^n.\eqno(1.13)$$
One should stress, however, that the representation of the type (1.11)
is preferable to that of (1.13). This fact was first noticed by F.
Coester and R. Haag [5]. I did systematically use the
 exponential representation to consider the boson models $g(\phi ^4)_2,
g(\phi ^4)_3$, and $g((\phi^*\phi)^2)_2$ [6]. Later I have generalized the
formalism to enable one to consider also fermions [7].  This generalization
essentially boils down to substituting expression (1.11) of the ground
state wave function into the Schroedinger equation, multiplying this
equation by the operator $e^{-K}$ and using eq. (2.10a)  (  see  also
comments after\\
eq. (2.10a) ).

{\bf 1.4.}  The consideration of the present work is of any value only
if one believes that the cut-off Schroedinger
equation (1.12) governs the QED.
Of course, the cut-off Schroedinger equation  approach to QED,
even if the Hamiltonian is bounded from below ( item 1.1.2.)
, gives rise to problems with the Lorentz invariance -cf., e.g., analogous
approach to the $g(\phi ^4)_4$ model. {\em I hope, these problemes can be
solved via a properly chosen realistic regularization (cut-off)}.
I do mean the introduction into the Fourier representation of the function
${\bf B}^{tr}({\bf x})$, eq.(1.6), of a photon
form-factor $F_{ph}({\bf k},l)$ and introduction
into the
analogous representation of the fermion operator $\psi _1({\bf x})$
of a fermion form-factor $F_f({\bf p},l)$. These cut-off representations
of the the vector potential and the fermion operators are to be used
only in the interaction terms $H_1, H_c$ and $H_p$.

{\bf 1.4.1.} Let us denote by $m_n^2$ the n-th order perturbation theory
contribution to the squared fermion mass in the spinor QED. Obviously,
one has $m_0^2=m^2$. Using the form -factors
$$
F_{ph}({\bf k},l)=\sum_{n\geq 0} F_{nph}e^{-n|{\bf k}|/l},
\sum_{n\geq 0} F_{nph}=1,
$$
$$
F_f({\bf p},l)=\sum_{n\geq 0} F_{nf}e^{-nE({\bf p})/l},
\sum_{n\geq 0} F_{nf}=1
,\eqno(1.14)
$$
I was able to exhibit the momentum dependence of the quantity $m_2^2$:
$$
m_2^2=e^2(m^2\ln (l/m)const_1+{\bf p}^2const_2+o(1/l)).
\eqno(1.15)
$$
Here, $o(x)\ra 0$ as $x\ra 0$.
The analogous momentum dependence show the second and third order
perturbation theory contributions to the squared mass of the fundamental
particle in the $g\phi ^4_4$ model.
I hope, it is possible to eliminate the momentum dependence of
the squared mass by a proper choice of the constants $F_n$ in
formfactors and thus, to construct the Lorentz-invariant
perturbation theory.

{\bf 1.4.2.} The textbook by W. Heitler [3] contains the calculation of
the quantity $m^2_{2F}$ which is the Feynman perturbation theory contribution
to the quantity $m^2_2$ ([6], Chapter 6, sec. 29, item 1., equation (29.14')).
The value of $m^2_{2F}$ does not depend on $|{\bf p}|$.

{\bf 1.4.3.} The point, however, is that $m^2_2-m^2_{2F}\e \d m^2_2\neq 0.$
The Hamiltonian $H_{QED}$ without the Palumbo term $H_p$ gives
$m^2_2=m^2_{2tr}+m^2_{2c}$, where subscripts "tr" and "c" denote parts of the
quantity $m^2_2$ which originate due to the
exchange of transversal photons and
due to the Coulomb interaction. As for the quantity $m^2_{2F}$, it can be
represented as $$m^2_{2F}=\sum_{{\bf k},\mu }
 Trace(A({\bf k,p})\g _{\mu }B({\bf k,p})\g _{\mu })
=\sum _{1}^{4}m^2_{2F\mu }$$
$$
=m^2_{2Ftr}+m^2_{2Flong}+m^2_{2F4},
$$
where $m^2_{2Ftr}=m^2_{2tr}$. Thus, one has $\d m^2_2=m^2_{2c}-m^2_{2Flong}
-m^2_{2F4}$. Straightfoward calculation gives
$$
\d m^2_2=e^2\int F_f({\bf p+k} ,l)F_{ph}({\bf k},l)
F_f({\bf p},l){\bf pk}/|{\bf k}|^3d{\bf k},
$$
 Using equations
$$|{\bf p+k}| =|{\bf k}|+{\bf pk}/|{\bf k}| +...,
F_f({\bf k},l)\ra \Phi _f(|{\bf k}|/l), F_{ph}({\bf k},l)
\ra \Phi _{ph}(|{\bf k}|/l)$$
as $l\ra \infty $, here, $\Phi _f(z)$ and $\Phi _{ph}(z)$ are some
functions, one gets
$$
\d m^2_2=e^2{\bf p}^2/3\int _{0}^{\infty }\Phi _{ph}(z)
((d/dz)\Phi _f (z))dz
$$
as $l\ra \infty $, cf. eq. (1.15) .

{\bf 1.4.4.} If the integral here equals zero, the second order
perturbation theory consideration is compatible with the Lorentz
invariance.

{\bf 1.5.} In principle, the term  $O(1/V)$ in eq. (2.23) is able to
reverse the result of my consideration. Let it be, e.g.,\\
$O(1/V)=const(\sum _{\l =1,2} q({-\bf r},\l )q({\bf r},\l ))^2/V$
and $const>0$. Then, the operator (2.23)
will be bounded from below so that my consideration
cannot exclude possibility that the Hamiltonian $H_{QED}$ possesses the
ground state. Let us denote it by $\O_{0V}$.
 Let us also denote by $\O_{0good}$
the ground state of the spinor QED, which it would possess,
were the operator (2.23) without the
term $O(1/V)$  be bounded from below. {\em The point
is that
these two vacua are as drastically different as for instance
the ground states of the quantum mechanical Hamiltonians $H_{1,0}$
and $H_{-1,1/V} , V\ra +\infty $, where $H_{a,b}=-(d/dz)^2+az^2+bz^4$.}

{\bf 1.6.} Eqation (0.3) and (0.4) (if $e^2\ll 1$)
testify to the existence of the
squared photon mass of the order of magnitude $\sim l^2$. The $e^2$
perturbation theory contribution to this quantity is negative in the
spinor QED and is positive in the scalar QED.

\vspace{5mm}

 {\bf 2.THE PROOF OF THE STATEMENT OF THE ABSTRACT}\\
\vspace{5mm}\\

I shall prove this statement in several steps.

 {\bf 2.} At first, I shall average the Hamiltonian (1.1) over the
 normalized photon state $\Omega _{ph}$,
 $$\Omega _{ph}=const\exp (-\omega[{\bf q}(0)^2+
\sum _{{\bf k}\neq 0,{\bf k}\neq \pm {\bf r};\lambda =1,2}
 q({\bf k},\lambda )q(-{\bf k},\lambda )]) ,
r\e |{\bf r}|, r,\omega >0
 , \eqno(2.1) $$
 i.e., I shall consider the transformation
 $$H_{QED}\rightarrow H_{QED1}= \int\Omega _{ph}^*H_{QED}\Omega _{ph}
 dS({\bf r}), \eqno(2.2)$$
see eq. (0.3a),
and analogous transformation $h\ra h_1$.
 Then, one gets  $H_p \rightarrow const,
 H_{0f} \rightarrow H_{0f},
 H_c \rightarrow H_c, $
 $$
H_{0ph}\ra H_{0ph1}\e \sum_{\l =1,2}[- \frac {\partial }{\partial
 q({\bf r},\l )}\frac{\p }{\p q(-{\bf r},\l )}+
r^2q({\bf r},\l )q(-{\bf r},\l )]+const,
\eqno(2.3a)
$$
$$
H_1\ra H_{11}\e \frac{e}{\sqrt{V}}
\sum _{{\bf p,s};{\bf s}=\pm {\bf r};\l ;\s ;\t }q({\bf s},\l )
[a^*({\bf p+s},\s )b^*({\bf-p},\t ){\bf A(p+s,p;\s ,\t )}
$$
$$
+b({\bf -p-s,\s })a({\bf p,\t }){\bf D(p+s,p;\s ,\t )}
$$
$$
+a^*({\bf p+s,\s })a({\bf p,\t }){\bf B(p+s,p;\s ,\t )}
$$
$$
+b({\bf -p-s,\s })b^*({\bf -p,\t }){\bf C(p+s,p;\s ,\t )}]\cdot {\bf e(s,\l )}
$$
$$
\e H_{11}(a^*b^*)+H_{11}(ba)+H_{11}(a^*a+bb^*)
, \eqno(2.3b)
$$
$$
H_2 \rightarrow  M^2(r^2\sum _{\l }q({\bf r,\l })q({\bf -r,\l })
+ const)\e H_{21}, \eqno(2.4)$$
 $$H_{QED1} = H_{0f} +H_{0ph1}+H_{11}+H_c  +const, \eqno(2.5)$$
 Here
 $${\bf A(p+s,p;\sigma ,\tau )}= u^*({\bf p+s},\sigma )
{\hat \alpha }v({\bf p},\tau ), $$
  $$ {\bf B(p+s,p;\sigma ,\tau )}=
u^*({\bf p+s},\sigma ){\hat \alpha }u({\bf p},\tau ),
  $$
  $$
  {\bf C(p+s,p;\s ,\t)}=v^*({\bf p+s},\s ){\hat \a }v{\bf (p,\t )}
  $$
  $$
  {\bf D(p+s,p;\s ,\t )}=v^*({\bf p+s},\s ){\hat \a }u({\bf p},\t ).
\eqno(2.6)
$$

 {\bf 2.1} Then, let us determine the function $\Omega _f$ and
the operator $K$,
 $$\Omega _f =e^K|0>,\eqno(2.7)$$
 $$K =\sum_{{\bf p,s,\s \t;s=\pm r} }
 K({\bf p+s,p};\sigma ,\tau )a^*({\bf p+s,\sigma })b^*(-{\bf p},\tau )
\e \sum K({\bf p+s,p}),\eqno(2.8)$$
 (here $|0>$ is the state of the fermion bare vacuum:
 $a({\bf p,\sigma })|0>=
 b({\bf p,\sigma })|0>=0$ for all values of ${\bf p}$ and $\sigma $) by the
equation
 $$(H_{0f}+H_{11}(a^*b^*))\O _f=0.\eqno(2.9)$$
 One easily gets
 $$K({\bf p+s,p};\s,\t)=-\sum_{\l }\frac
 {e q({\bf s,\l }){\bf A(p+s,p;\s,\t)}\cdot {\bf e(s,\l )}}
  {\sqrt{V}(E(p)+E(|{\bf p+s}|))}.
 \eqno(2.10)$$
In order to derive eq. (2.10) from eq. (2.9), it is sufficient to multiply
eq.(2.9) by the operator $e^{-K}$ and apply the formula
$$
e^{-K}Ae^{K}=A+[A,K]+\h [[A,K],K]+...
\eqno(2.10a)
$$
(where square brackets denote the commutator), to the operators
$A_1\e H_{0f}$ and $A_2\e H_{11}(a^*b^*)$ . For the second operator all the
commutators in eq. (2.10a) disappear, analogously for the  operator $A_1$
the decomposition in the r.h.s. of eq. (2.10a) reduces to its
first two terms. Thus, equation (2.9) becomes trivial.\\
(Note that if the operator $A$ were for instance bilinear in annihilation
operators and not contain derivatives with respect to boson variables,
the series (2.10a) would reduce to its first three terms.)\\
It follows from eqs. (2.7) and (2.8) that
$$\O_f=\prod _{{\bf p}}(1+\sum _{{\bf s=\pm r}}K({\bf p+s,p})
+\h (\sum _{{\bf s=\pm r}}K({\bf p+s,p}))^2)|0>.\eqno(2.11)$$
Here, $\prod _{{\bf p}}$
denotes the product over all values of ${\bf p}$, $|{\bf p}|<l$,
terms with $|{\bf p+s}|>l$ have to be omitted.

Let us denote the quantity $\O^*_f\O_f$ by $Q$. Eqs. (2.8) and (2.11) give
$$Q=(\prod_{{\bf p}}(1+\sum _{{\bf s=\pm r}}D_1({\bf p+s,p})
+\sum _{{\bf s=\pm r}}D_2({\bf p+s,p})+D_3({\bf p,r})),
$$
$$
D_1({\bf p+s,p})=<0|^*K({\bf p+s,p})^*K({\bf p+s,p})|0>,
$$
$$D_2({\bf p+s,p})= <0|^*(K({\bf p+s,p})^*)^2K({\bf p+s,p})^2|0>/4,
$$
$$
D_3({\bf p,r})=<0|^*K({\bf p+r,p})^*K({\bf p-r,p})^*K({\bf p-r,p})
K({\bf p+r,p})|0>,
$$
$$ D_1({\bf p+s,p})=O(1/V),
D_2({\bf p+s,p})=O(1/V^2), D_3({\bf p,r})=O(1/V^2).\eqno(2.12)$$
We shall introduce the quantities $Q_1$ and $D_1$,
$$Q_1=e^{D_1}, D_1=\sum_{{\bf p,s;s=\pm r}} D_1({\bf p+s,p}).\eqno(2.13)$$

{\bf 2.1.1.} The following important formula holds:
$$Q=Q_1(1+O(1/V)).\eqno(2.14)$$

{\bf 2.1.2.} Equations (2.8) and (2.10) result in the definition
$$K({\bf p+s,p})\e \sum_{\l } q({\bf s,\l })K({\bf p+s,p},\l ).\eqno(2.15)$$
Here the function $K({\bf p+s,p},\l )$ does not depend on the vector
potential variables ${\bf q}(0), q({\bf k,\l })$.
Eqs. (2.6), (2.9), (2.10), (2.14) and (2.15) give
$$
\O^*_f(H_{0f}+H_{11})\O_f=\O^*_f(H_{11}(ba)+H_{11}(a^*a+bb^*))\O_f\e
Z_1+Z_2,\eqno(2.16)$$
$$Z_1=-Qe^2\sum_{\l }q({\bf r,\l })q({\bf -r,\l })Z(r,m,l)
+O(1/V)),\eqno(2.17)$$
$$
Z(r,m,l)=\frac{2}{(2\pi )^3}\int_{|{\bf p}|<l,|{\bf p+r}|<l}\frac
{E({\bf p+r})E({\bf p})+({\bf pr})^2/r^2+{\bf pr}-m^2}
{E({\bf p})E({\bf p+r})(E({\bf p})+E({\bf p+r}))}d{\bf p}.
\eqno(2.17a)
$$
The quantity $Z_2$ evidently , equals zero:
$$
Z_2=0.\eqno(2.18)
$$
Let us introduce the notation $\O_{f1}$:
$$\O_{f1}\e \O_f/\sqrt{Q},\quad \O_{f1}^*\O_{f1}=1.\eqno(2.19)$$
There hold  the formulas (see Appendix A)
$$\O_{f1}^*\sum_{\l }\frac{\p }{\p  q({\bf r,\l })}
\frac{\p }{\p q({\bf -r,\l })}\O_{f1}=
\sum_{\l }[\frac{\p }{\p q({\bf r,\l })}\frac{\p }{\p q({\bf -r,\l })}
$$
$$
+\sum_{{\bf s=\pm r}} X({\bf s,\l })
\frac{\p }{\p q({\bf s,\l })}]+Y,
\eqno(2.20)$$
$$ X({\bf s,\l })=O(1/V), Y=const+O(1/V).\eqno(2.20a)$$

{\bf 2.1.3.} Now let us consider the quantity $C$,
$$C=\O_{f1}^*H_c\O_{f1}.\eqno(2.21)$$
It is convenient to represent $H_c$ in a normal form. We shall symbolically
write down this representation as $H_c=const_1 + const_2(a^*a+b^*b)
+const_3(a^*b^*+
ba) + const_4(a^*a^*b^*b^*+a^*b^*ba+bbaa)$. Correspondingly, we shall
represent $C$ as $C=C_1+C_2+C_3+C_4 $.
Then, $C_1$ does not depend on the
variables $ q({\bf s,\l ),s=\pm r},$, while
$C_2$ and $C_4$ depend on these variables
quadratically and $C_3=0$. Rotational invariance and
dimensional considerations give
$$C=e^2mf(m/l,r/l)- \sum_{\l }q({\bf r,\l })q({\bf -r,\l })e^4d(m/l,r/l)l^2
+O(1/V).\eqno(2.22)$$
Here,  $f(x,y)$
 and $d(x,y)$
are some functions.

{\bf 2.2.} Equations (2.3a), (2.3b), (2.5)  and (2.16-22) prove the formula
$$\O_{f1}^*H_{QED1}\O_{f1}=-\sum_{\l =1,2}[\frac{\p }{\p q({\bf r,\l })}
\frac{\p}{\p q({\bf -r,\l })}+
$$
$$ q({\bf r,\l })q({\bf -r,\l })(e^2c(m/l,r/l)
+e^4d(m/l,r/l))l^2]+const+O(1/V), c(x,0)>0,\eqno(2.23)$$
-cf. the formulas (0.3) and (0.4). Equation (2.4) gives
$$\O_{f1}^*H_{2,1}\O_{f1}= M^2(\sum_{\l =1,2}q({\bf r,\l })
q({\bf -r,\l })+const).$$
Last two formulas complete the task of this section.

{\bf 2.2.1.} The starting point of my consideration of the problem of
unboundedness from below of the operators (2.23) and the like is
the statement that the operator $-(d/dz)^2-\g ^2z^2, \g ^2>0,$
is unbounded from below.
\vspace{5mm}\\

\vspace{5mm}
ACKNOWLEDGEMENT

I am  deeply indebted to Professor M.Consoli whose remark enabled me
to get at the construction of eq. (2.10)
and thus to overcome the impass
I was in. I am obliged very much also to Dr. Ch.Devchand for his kind
interest in the work and to Drs. A.B.Govorkov and M.I.Shirokov
for useful criticism and remarks.

\vspace{5mm}

APPENDIX A\\
\vspace{5mm}\\
Here I shall prove eq. (2.20a). Equations (2.19) and (2.20) give
$$ X({\bf s,\l })=\O_f^*\sum_{{\bf p}} K({\bf p+s,p},\l )\O_f/Q+\sqrt{Q}
\frac{\p }{\p q({\bf s,\l })}(1/\sqrt{Q}),
\eqno(A1)$$
$$Y=\O_f^*\sum_{{\bf p}1,{\bf p}2,\l } K({\bf p}1+{\bf r,p}1,\l )
K({\bf p}2-{\bf r,p}2,\l )\O_f/Q
$$
$$+
\O_f^*\sum_{{\bf p,s,\l ;s=\pm r}} K({\bf p-s,p},\l ))\O_f/\sqrt{Q}
\frac{\p }{\p  q({\bf s,\l })}(1/\sqrt{Q})
$$
$$
+\sum_{\l }\sqrt{Q}\frac{\p }{\p q({\bf r,\l })}
\frac{\p }{\p q({\bf -r,\l })}
(1/\sqrt{Q})\e Y_1+Y_2+Y_3.\eqno(A2)$$
Equations  (2.11)- (2.14) give
$$\O_f^*\sum_{{\bf p}} K({\bf p+s,p},\l )\O_f=
$$
$$
=\sum_{{\bf p}} <0|^*K({\bf p+s,p})^*K({\bf p+s,p},\l )|0>Q(1+O(1/V))
$$
$$
=\h \frac{\p }{\p q({\bf s,\l })}Q_1
[1+O(1/V)],\eqno(A3)$$
$$Y_1=\sum_{{\bf p}1,{\bf p}2,\l }
 <0|^*K({\bf p}1+{\bf r,p}1)^*K({\bf p}1+{\bf r,p}1,\l)|0>
$$
$$
<0|^*K({\bf p}2-{\bf r,p}2)^*K({\bf p}2-{\bf r,p}2,\l )|0>(1
+O(1/V))
$$
$$
=(\h ) ^2Q_1^{-2}\sum_{\l }\frac{\p Q_1}{\p q({\bf r,\l })}
\frac{\p Q_1}{\p q({\bf -r,\l })}(1 +O(1/V)).
\eqno(A4)$$
It follows from equations (A2) and (A3) that
$$Y_2=\h \sum_{{\bf s=\pm r;\l }}(\sqrt{Q}\frac{\p }
{\p q({\bf s,\l })}[1/\sqrt{Q}])
(\frac{\p }{\p q({\bf -s,\l })}Q_1)/Q_1(1+O(1/V)).\eqno(A5)
$$
Now note that eqs. (2.10)-(2.15) result in the formula
$$Q=\exp(const(\sum_{\l =1,2} q({\bf r,\l })q({\bf -r,\l })))(1+O(1/V).
\eqno(A6)$$
So, equations (A1)-(A6) and eq. (2.14) entail eq. (2.20a).
This result completes the consideration of  Appendix A.
\vspace{5mm}\\

References\\
1.Palumbo F. Phys. Lett.{\bf B173} (1986) 81.\\
2.L\" {u}sher M. Nucl. Phys.{\bf B219} (1983) 233.\\
3.Heitler W. Quantum Theory of Radiation, Oxford, Clarendon Press, 1954.\\
4.Zastavenko L.G. Preprint JINR E2-90-280 (1990),unpublished. \\
5.Coester F., Haag R.,Phys. Rev.{\bf 117} (1960) 1137.\\
6.Zastavenko L.G. TMF {\bf 7} (1971) 20, TMF {\bf 8} (1971) 335,
 TMF {\bf 9} (1971) 355.\\
7.Zastavenko L.G. Preprint JINR E-2-7725 (1974), unpublished.\\

\end{document}